\documentclass[10pt, prd, twocolumn]{revtex4}

\usepackage{amsmath}
\usepackage{subfigure}
\usepackage{amssymb}
\usepackage{graphicx}
\usepackage{bm} 
\usepackage{xcolor}

\newcommand{\be}[0]{\beta}

\def \be {\begin{equation}}
\def \ee {\end{equation}}
\def \ba {\begin{aligned}}
\def \ea {\end{aligned}}
\def \bea {\begin{eqnarray}}
\def \eea {\end{eqnarray}}

\begin{document}

\title{Path integral and instantons for the dynamical process and phase transition rate\\
 of Reissner-Nordstr\"om-AdS black holes}

\author{Conghua Liu$^{a,b}$, Jin Wang$^{c,}$\footnote{Corresponding author.\\jin.wang.1@stonybrook.edu} }
\affiliation{$^{a}$College of Physics, Jilin University, Changchun 130022, China\\
$^{b}$State Key Laboratory of Electroanalytical Chemistry, Changchun Institute of Applied Chemistry, Chinese Academy of Sciences, Changchun 130022, China\\
 $^{c}$Department of Chemistry and Department of Physics and Astronomy, State University of New York at Stony Brook, Stony Brook, New York 11794, U.S.A.}

\begin{abstract}

We propose a new approach to study the phase transition dynamics of the RNAdS black holes on the underlying free energy landscape. By formulating a path integral framework, we quantify the kinetic paths representing the histories from the initial state to the end state, which provides us a visualized yet quantified picture about how the phase transition proceeds. Based on these paths, we derive the analytical formulas for the time evolution of the transition probability and provide a physical interpretation of the contribution to the probability from one ``pseudomolecule" (``anti-pseudomolecule"), composed of instantons and anti-instantons, which is actually the phase transition rate from the small(large) to the large(small) black hole state. These numerical results show a good consistency with the underlying free energy landscape topography.

\end{abstract}


\maketitle



\newpage

\section{Introduction}

In general relativity, the classical black holes are emerged from the solutions of Einstein's equation. The black holes have some fascinating features. They are perfect absorbers but emit nothing. As known, an object with non-zero temperature has the thermal radiation. This implies that the physical temperature of the classical black hole is zero and the black hole thermodynamics seems to be impossible. This has all been changed since the appearance of the black hole area law~\cite{AA}, stating that the event horizon area of the black hole can never decrease with the time. Bekenstein incisively noticed the similarities with the second law of thermodynamics, and proposed that every black hole should have its own entropy which is associated with the event horizon area by a directly proportional relationship~\cite{AB}. Thereafter, the four laws of black hole mechanics~\cite{AC} were formulated, analogous to the four laws of thermodynamics. However, since the temperature of the classical black hole is zero, it implies that these similarities are merely formal and do not have profound physical implications.

The whole picture has been significantly altered since the quantum effects were considered, leading to the famous Hawking radiation, which shows that the black holes emit radiation with a blackbody spectrum~\cite{AD}. The Hawking radiation and the four laws of black hole mechanics indicate that the black holes are thermodynamic systems with temperatures. Since then, the black hole thermodynamics has been widely used to study the black hole physics. A famous example is the Hawking-Page phase transition occurring in the asymptotically anti-de Sitter space, where a first-order phase transition has been found between the thermal radiation and the large stable Schwarzschild anti-de Sitter black hole~\cite{AE}. The recent researches show that there is a correspondence between the gravitational physics in anti-de Sitter space (AdS) in the bulk and one dimensional less conformal field theories (CFT) on the boundary via holography~\cite{AF,DA,DB,CA,CB}. In the context of AdS/CFT correspondence, the Hawking-Page transition can be interpreted as the confinement/deconfinement phase transition in QCD~\cite{AF}. By viewing the cosmological constant as the thermodynamic pressure in the AdS space, the analogues of the charged-AdS black holes and the Van der Waals fluids have been explored~\cite{AH,AI,AG,CA,CB}. The behaviors of the black hole thermodynamics at the triple point phase transition have also been investigated~\cite{AK}. All of these studies provide us a more profound understanding to black hole thermodynamics.

However, the dynamics of how a black hole state or phase transforms to another one during the phase transition has not been investigated adequately until very recently. The very recent studies of black hole phase transition dynamics under thermal fluctuations have been explored on the free energy landscape by solving the corresponding probabilistic Fokker-Planck equation, giving rise to the fluctuations and the mean first passage time (i.e. the inverse of the rate)~\cite{AL,AM}. The complete description of the dynamics should include two aspects. The rate shows how fast the black hole phase transition occurs and the path shows how the process proceeds in the phase transition. Thus, it is necessary for us to quantify the phase transition path to explore the underlying dynamical process.

Since its appearance in~\cite{AN}, the path integral methods have been developed and used to study many physical and chemical problems successfully~\cite{AO,AP,AQ,AR,AS}. The advantage of such method is that one can quantify the paths with weights representing the histories from the initial state to the end state. The paths will provide us a quantitative and visual picture of the phase transition process. This certainly helps us to understand the dynamics of phase transition better.

The phase transition of RNAdS black holes takes place in the asymptotically AdS space with the negative cosmological constant. By interpreting the cosmological constant as thermodynamic pressure~\cite{AH,AT,AU}, one can formulate the extended phase space and study the Van der Waals type phase transition in RNAdS black holes. By choosing the black hole radius as the order parameter, the free energy landscape can be quantified along this order parameter. The phase transition can then be easily analyzed on the free energy landscape~\cite{AL,AM}. There are three macroscopic emergent phases, the small, the thermodynamic transition and the large black hole states. The small and large black hole states are locally stable and the thermodynamic transition black hole state is unstable. Under the thermodynamic fluctuations, the phase transition is possible between the locally stable small black hole state and the locally stable large black hole state.

In this paper, we study the process of such a phase transition by using a path integral method~\cite{AR,AS,BA,BB,BC,BD,BE,BF,AV}. The weights of different paths are from the exponentials of the path integral actions. This implies that the weight of the dominant path is significantly larger than that of the other paths due to the exponential suppression in the weights of the other paths. Then we can just consider the contribution from the dominant path. The dominant path should satisfy the Euler-Lagrangian equation due to the minimization of the action or maximization of the weight, and we can transform the E-L equation as an energy conservation equation. Thus, the phase transition between the small and the large black hole states can be regarded as a one-dimensional particle under an effective potential moving between the corresponding small and large black hole states.  In the long time limit, the phase transition can go back and forth many times and the kinetic paths can be quantified. The dominant path is composed of a series of smallest units named pseudomolecules, with each made of a pair of instantons (we have referred ~\cite{BD,BF,AV} for using the words ``pseudomolecule" and ``instanton"). By assuming that there are no interactions between the instantons, we can quantify the probability in the dilute gas approximation. We find that the contribution to the probability from one pseudomolecule or one anti-pseudomolecule is actually responsible for quantifying the phase transition rate from the small to large black hole state or from the large to small black hole state. The expressions of the phase transition rates can be obtained analytically. All these results are consistent with the underlying Gibbs free energy landscape. This paper presents a new framework to study the dynamical phase transition process of the black holes and the spacetimes. We address the crucial kinetic path issue and provide a more profound understanding to the phase transition process of the RNAdS black holes.

The paper is organized as follows. In sec.~\ref{ToRNAdS}, we illustrate the thermodynamic properties of RNAdS black holes under the underlying free energy landscape. In sec.~\ref{PIaPT}, we introduce the path integral framework and apply it to the phase transition of RNAdS black holes. Then, the kinetic paths, phase transition rates, and the time evolutions of the probabilities are presented. In sec.~\ref{Conclusions}, we present the conclusions.

\section{Thermodynamics of RNAdS black hole and the free energy landscape}
\label{ToRNAdS}

In this section, we will briefly review the thermodynamic properties of RNAdS black holes~\cite{AH,AL,AM,BI,BJ}.

The metric of RNAdS black hole is given by ($G=1$ units)
\be
ds^2=-f(R)dt^2+\frac{dR^2}{f(R)}+R^2d\Omega^2,
\ee
where $f(R)$ is given by
\be
f(R)=1-\frac{2m}{R}+\frac{q^2}{R^2}+\frac{R^2}{l^2}.
\ee
The parameter $m$ represents the black hole mass, $q$ is the black hole charge, and $l$ is the AdS curvature radius which is associated with the negative cosmological constant $\Lambda$ by $l=\sqrt{\frac{-3}{\Lambda}}$.

In the AdS space, the cosmological constant can be interpreted as the thermodynamic pressure in extended thermodynamics~\cite{AH,CC,CD,CE}:
\be
\label{eq:2.5}
P=\frac{3}{8\pi}\frac{1}{l^2}.
\ee

The Hawking temperature is given by

\be
\label{eq:2.7}
T_H=\frac{1}{4{\pi}r}(1+8{\pi}Pr^2-\frac{q^2}{r^2}).
\ee

We should note that there is a critical pressure $P_c=\frac{1}{96{\pi}q^2}$~\cite{AH,AM}. When $P>P_c$, the Hawking temperature $T_H$ is a monotonic increasing function of $r$. When $P<P_c$, $T_H$ has a local minimum value $T_{min}$ and a local maximum value $T_{max}$, which are determined by $\frac{{\partial}T_H}{{\partial}r}=0$. We will focus on the regime $P<P_c$ and $T_{min}<T<T_{max}$, where there are three on-shell solutions to the stationary Einstein field equation as the small, the thermodynamic transition and the large black holes.

On the free energy landscape, the free energy of the system is defined as a continuous function of the order parameter. It is necessary to introduce a series of off-shell states for the study of the black hole phase transition dynamics. In general, we can choose the radius of the AdS black hole as the order parameter and assume a  canonical ensemble which is composed of various black hole spacetime states with different radii at the specific temperature~\cite{AE,AL,AM,BI,BJ}. This includes all the possible states appearing during the phase transition. These states are characterized by the different black hole radii. Except for the small, the thermodynamic transition and the large black holes, all the other states are off-shell and do not obey the stationary Einstein field equation. In fact, the off-shell black hole states are unstable transients resulting from the classical thermal fluctuations of the black hole spacetimes. They are important in characterizing the free energy landscape of the black hole space time instead of the isolated stable black hole phases. Although the off-shell states are intermediate transient and unstable, they are still significant as a bridge connecting between the stable black hole phases reflected by the dominant phase transition paths as illustrated in Fig.~\ref{fgIP}, where the associated dynamics of the phase transition is explicitly shown (This will be illustrated in detail in the next section). In the recent research, it was found that there is a lower bound for the order parameter which corresponds to the extremal black hole~\cite{CF}. We denote the lower bound of the order parameter as $r_{ex}$.

Replacing the Hawking temperature $T_H$ by the ensemble temperature $T$ in the on-shell Gibbs free energy expression $G=m-T_HS$, we can generalize the on-shell Gibbs free energy to the off-shell free energy as~\cite{AM,BI,BJ}:
\be
\label{eq:2.9}
G=m-TS=\frac{r}{2}(1+\frac{8}{3}{\pi}Pr^2+\frac{q^2}{r^2})-{\pi}Tr^2,
\ee
where the order parameter $r$ can take the continuous values from $r_{ex}$ to infinity.

\begin{figure}[t]
\centering
\includegraphics[width=0.48\textwidth]{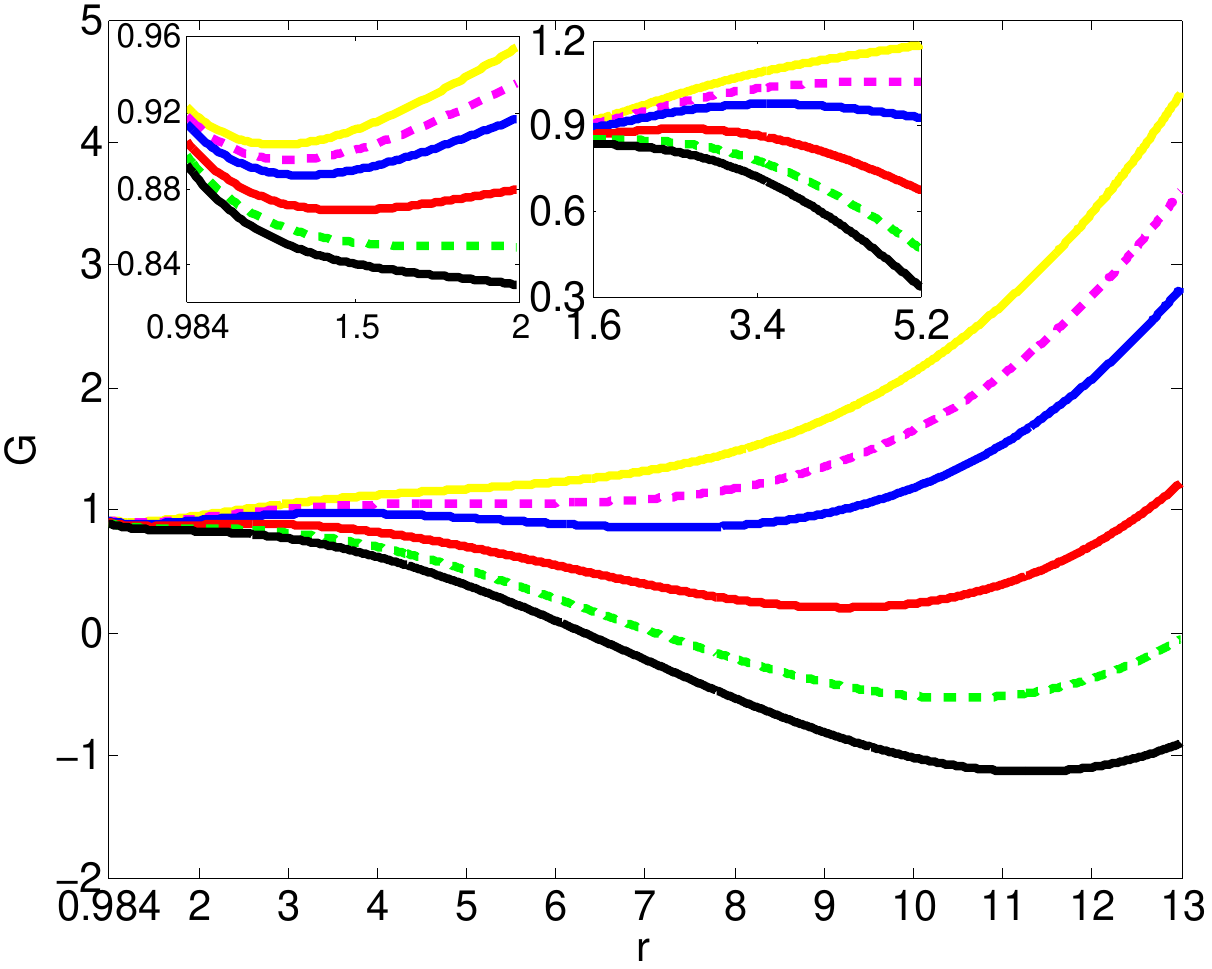}
\caption{The Gibbs free energy as the function of black hole radius $r$ at different temperatures. The values of the temperature from top to bottom are set as: $T=0.027<T_{min}$, $T=T_{min}=0.0285$, $T_{min}<T=0.03<T_{max}$, $T_{min}<T=0.033<T_{max}$, $T=T_{max}=0.0354$, $T=0.037>T_{max}$.}
\label{fgFE}
\end{figure}

We choose $P=0.4P_c$ and $q=1$ in all the next calculations, $r_{ex}$ can be calculated as 0.984, which is smaller than the radius of the small black hole at various temperatures. In Fig.~\ref{fgFE}, we have plotted the free energy as a function of black hole radius $r$ at different temperatures.
From the figure, it can be seen that when $T_{min}<T<T_{max}$, the Gibbs free energy has three local extremum points (a local maximum point and two local minimum points). They satisfy the equation:
\be
\label{eq:2.10}
\frac{\partial G}{\partial r}=\frac{1}{2}+4{\pi}Pr^2-\frac{q^2}{2r^2}-2{\pi}Tr=0,
\ee
which is same as eq.~(\ref{eq:2.7}) when we replace the Hawking temperature $T_H$ by the ensemble temperature $T$. Therefore, the radii of the three local extremal points are actually the three on-shell solutions to the stationary Einstein field equation.

Based on the condition of $P=0.4P_c$ and $q=1$, we can solve eq.(\ref{eq:2.10}) and obtain the values of $r_s$, $r_m$ and $r_l$ at different temperatures.

Furthermore, it can be seen from the Fig.~\ref{fgFE}, the small and large black hole states corresponding to the free energy minima are locally stable while the thermodynamic transition black hole state corresponding to the free energy maximum is unstable. Because of the thermal fluctuations, it is possible for the phase transitions between the locally stable small black hole state and the locally stable large black hole state. We will study the dynamics of such phase transitions by using path integral methods in the next section.

\section{Path-integral and phase transition rate of RNAdS black hole}
\label{PIaPT}
\subsection{The path integral of the RNAdS black hole phase transition}

The stochastic dynamics of the RNAdS black hole under the thermal fluctuations can be described by the probabilistic Fokker-Planck equation in{~\cite{AM}}. In order to formulate the path integral framework, it is convenient to use the equivalent stochastic Langevin equation for the trajectories as:
\be
\ba
\label{eq:3.2}
\frac{d r}{d t}=-\frac{\partial G(r)}{\gamma\partial r}+\eta(r,t),
\ea
\ee
where $\gamma$ is the friction coefficient; $-\frac{\partial G(r)}{\gamma\partial r}$ is the driving force; $\eta(r,t)$ is the fluctuating stochastic force. We assume that $\eta(r,t)$ is the Gaussian white noise, which satisfies the equations $\langle\eta(r,t)\rangle=0$ and $\langle\eta(r,t)\eta(r,0)\rangle=2D\delta(t)$. The diffusion coefficient $D$ is associated with the friction coefficient by the Einstein relationship
\be
D\gamma=k_{B}T.
\ee

The weight or the probability from the initial to final state at time t can be quantified by the Onsager-Machlup functional path integral as~\cite{AO,BA}:
\be
\ba
\label{eq:3.1}
P(r_t,t,r_0,0)=&\int Dr \exp\{-\int L[r(t)]dt\}\\
=&\int Dr \exp\{-\int[\frac{1}{4}\frac{(\frac{dr}{dt}+\frac{D\partial \beta G(r)}{\partial r})^2}{D}\\
&-\frac{1}{2}\frac{\partial(D\frac{\partial\beta G(r)}{\partial r})}{\partial r}]dt\},
\ea
\ee
where $Dr$ represents the sums of all the paths connecting the initial state and the end state, $L[r(t)]$ is the stochastic Lagrangian (also called the Onsager-Machlup functional):
\be
\ba
\label{eq:3.3}
L=\frac{1}{4}\frac{(\frac{dr}{dt}+\frac{D\partial \beta G(r)}{\partial r})^2}{D}-\frac{1}{2}\frac{\partial(D\frac{\partial\beta G(r)}{\partial r})}{\partial r}.
\ea
\ee

If we assume that the diffusion coefficient is very small, then the last term of the Lagrangian in eq.(\ref{eq:3.3}) can be ignored as:
\be
\label{eq:3.16}
L=\frac{1}{4}\frac{(\frac{dr}{dt}+\frac{D\partial\beta G(r)}{\partial r})^2}{D}.
\ee

\begin{figure*}[tbp]
\centering
\includegraphics[width=0.96\textwidth]{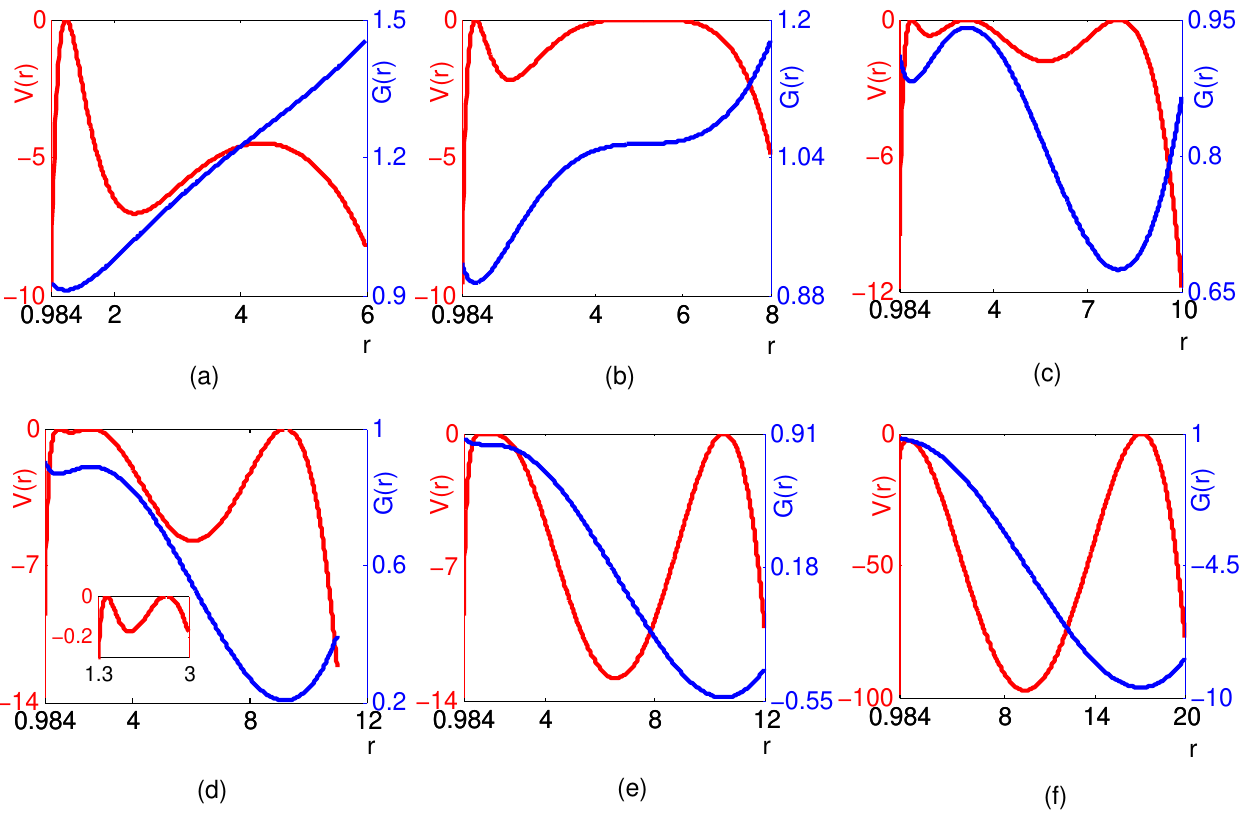}
\caption{The Gibbs free energy (blue line) and the effective potential (red line) as the functions of black hole radius at different temperatures: (a)$T=0.025<T_{min}$, (b)$T=T_{min}=0.0285$, (c)$T_{min}<T=0.031<T_{max}$, (d)$T_{min}<T=0.033<T_{max}$, (e)$T=0.0354=T_{max}$, and $(f)T=0.05>T_{max}$.}
\label{fgEV}
\end{figure*}

From Eq.(\ref{eq:3.1}), we can see that the different paths contribute to different weights, which are on the exponentials. This indicates that the dominant path has the largest weight, which can be significantly larger than the weights of the other paths due to the exponential suppression in the weight of the other paths. Thus, we can just consider the contributions of dominant path. The dominant path should obey the Euler-Lagrangian equation from the maximization of the weights or minimization of the action:
\be
\label{eq:3.17}
\frac{d}{dt}\frac{\partial L}{\partial\dot{r}}-\frac{\partial L}{\partial r}=0.
\ee

Substituting the Eq.(\ref{eq:3.16}) into the Eq.(\ref{eq:3.17}), we can obtain
\be
\label{eq:3.18}
\frac{d^2r}{dt^2}-\frac{1}{2}\frac{\frac{\partial D}{\partial r}}{D}\dot{r}^2-2D\frac{\partial u}{\partial r}=0,
\ee
where
\be
\label{eq:3.19}
u(r)=\frac{D}{4}(\frac{\partial\beta G(r)}{\partial r})^2.
\ee

Integrating the Eq.(\ref{eq:3.18}), we can obtain:
\be
\label{eq:3.20}
\frac{(\frac{dr}{dt})^2}{4D}-u(r)=E,
\ee
where E is a constant.

The eq.(\ref{eq:3.20}) can be regarded as an energy conservation equation. $\frac{1}{4D}(\frac{dr}{dt})^2$ is the kinetic energy term, $V(r)=-u(r)$ is the effective potential, and $E$ is the total energy. Thus, the dynamics of the phase transition can be described equivalently as the dynamics of one-dimensional particle with mass $\frac{1}{2D}$ moving in the effective potential $V(r)$~\cite{AP,AQ,AR,AS}.

Let us assume that $D$ is a constant and choose $D=k=1$ without loss of generality. In Fig.~\ref{fgEV}, we have plotted the Gibbs free energy and the effective potential as the functions of the black hole radius at different temperatures. It can be seen that when $T<T_{min}$ or $T>T_{max}$, there is only one state whose effective potential is zero. When $T=T_{min}$ or $T=T_{max}$, there are two such states. And when $T_{min}<T<T_{max}$, there are three such states. Analyzing eq.(\ref{eq:2.10}) and eq.(\ref{eq:3.19}), we can see all these radii of zero effective potential states are precisely the on-shell solutions to the stationary Einstein field equation (i.e. the extremum  points of the Gibbs free energy), which are also well shown in Fig.~\ref{fgEV}.

When $T_{min}<T<T_{max}$, there are three local maximums whose effective potentials are zero, representing the small, the thermodynamic transition and the large black hole states. Correspondingly, we denote their radii as $r_s$, $r_m$ and $r_l$. In the long time limit, the phase transitions between the small black hole state and the large black hole state can take place many times because of the thermodynamic fluctuations. This implies that the equivalent particle can go back and forth many times between the point $r=r_s$ and the point $r=r_l$ in the effective potential $V(r)$. The dominant path is composed of a series of smallest units of such jumps named pseudomolecules. These pseudomolecules should start at the locally stable states and end also at the locally stable states. Actually, every pseudomolecule is composed of a pair of instantons (or named pseudoparticles) whose paths are between the state of $r_s$ or $r_l$ and the state of $r_m$. There are four kinds of pseudomolecules in total: $a$ pseudomolecule has the trajectory $r_s\rightarrow r_m\rightarrow r_s$ with an instanton $r_s\rightarrow r_m$ and an anti-instanton $r_m\rightarrow r_s$, whose contribution to the probability is named $m_1$; $b$ pseudomolecule has the trajectory $r_s \rightarrow r_m \rightarrow r_l$ with a pair of instantons $r_s\rightarrow r_m$ and $r_m\rightarrow r_l$, whose contribution to the probability is named $m_2$; $c$ pseudomolecule has the trajectory $r_l\rightarrow r_m\rightarrow r_l$ with an anti-instanton $r_l\rightarrow r_m$ and an instanton $r_m\rightarrow r_l$, whose contribution to the probability is named $m_3$; $d$ pseudomolecule has the trajectory $r_l\rightarrow r_m\rightarrow r_s$ with a pair of anti-instantons $r_l\rightarrow r_m$ and $r_m\rightarrow r_s$, whose contribution to the probability is named $m_4$. We assume that there are no interactions between the instantons, so that we can calculate the final contribution by summing overall in the dilute gas approximation~\cite{BF,BH}.

\begin{figure*}[tbp]
\centering
\includegraphics[width=0.96\textwidth]{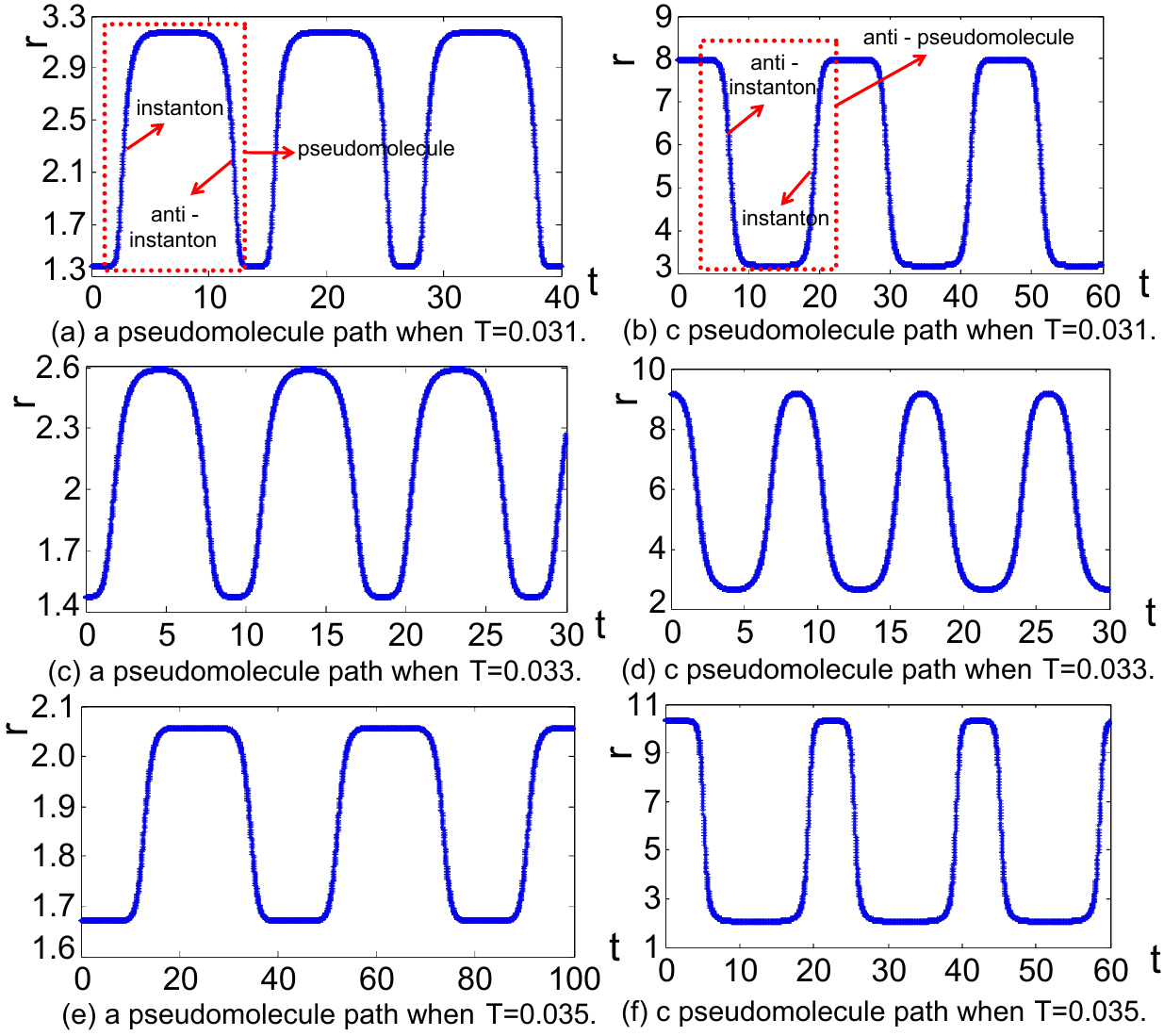}
\caption{The paths of $a$ and $c$ pseudomolecules and corresponding instantons and anti-instantons at different temperatures: $T=0.031$, $T=0.033$ and $T=0.035$.}
\label{fgIP}
\end{figure*}

In order to calculate the probability for the phase transition, we need to obtain the pseudomolecule paths. Based on the eq.(\ref{eq:3.18}), we can plot the black hole radius as the function of time $t$ from small (large) black hole state to the thermodynamic transition black hole state at different temperatures in Fig.~\ref{fgIP}, they are actually the paths of $a$ and $c$ pseudomolecules. After introducing the off-shell states, the dynamical process during the phase transition can be revealed clearly in the paths. From the paths, it can be seen that the phase transition between the small and large black holes will not have a residence time in the off-shell states. This indicates that the off-shell states are unstable transient states. 

The weight of one pseudomolecule contribution to the probability is given by:
\be
\label{eq:3.50}
M=\exp[-S]=\exp[-\int_{t_{initial}}^{t_{end}}L(r(t))dt],
\ee
where $S$ represents the action of the path:
\be
\ba
\label{eq:3.22}
S&=\int_{t_{initial}}^{t_{end}}L(r(t))dt\\
&=\frac{1}{4}\int_{t_{initial}}^{t_{end}}\frac{1}{D}(\frac{dr}{dt})^2+2\frac{dr}{dt}\frac{\partial\beta G(r)}{\partial r}+D(\frac{\partial\beta G(r)}{\partial r})^2dt.
\ea
\ee
For the weight contributions of the paths to the probability, we assume that the initial state is located at the locally stable small black hole state or the locally stable large black hole state whose effective potential and kinetic energy are zero. The energy conservation equation (\ref{eq:3.20}) becomes:
\be
\label{eq:3.51}
\frac{dr}{dt}=\pm\sqrt{4Du(r)}=\pm D|\frac{\partial\beta G(r)}{\partial r}|.
\ee
Based on the free energy figure in Fig.\ref{fgFE}, we know that the $\frac{\partial\beta G(r)}{\partial r}$ is greater than zero between $r_s$ and $r_m$, and less than zero between $r_m$ and $r_l$. The sign of $\frac{dr}{dt}$ is determined by the process in which the system proceeds. It is positive when the system translates from the small to large black hole state and negative from the large to small black hole state. Then, we can obtain:

From $r_s$ to $r_m$, $\frac{dr}{dt}=D\frac{\partial\beta G(r)}{\partial r}$;

From $r_m$ to $r_l$, $\frac{dr}{dt}=-D\frac{\partial\beta G(r)}{\partial r}$.

From $r_l$ to $r_m$, $\frac{dr}{dt}=D\frac{\partial\beta G(r)}{\partial r}$;

From $r_m$ to $r_s$, $\frac{dr}{dt}=-D\frac{\partial\beta G(r)}{\partial r}$.

Substituting these equations back into eq.(\ref{eq:3.22}), and being careful about the signs, we can see that the action $S=0$ for the trajectories from $r_m$ to $r_s$ and from $r_m$ to $r_l$. This indicates that only the trajectory from the small (large) black hole state to the thermodynamic transition black hole state contributes to the probability for the phase transition from small(large) black hole state to large (small) black hole state, which also corresponds to the free energy figures in Fig.~\ref{fgFE}. In the transition from the small to large black hole states, the free energy is uphill for the trajectory from the small black hole state to the thermodynamic transition black hole state and downhill from the thermodynamic transition black hole state to the large black hole state. The probability for the latter trajectory is 1, so we just need to consider the probability contribution for the former trajectory. The similar analysis can be made in the transition from the large black hole state to the small black hole state. Thus, in the phase transition from small (large) black hole state to large (small) black hole state, we do not need to obtain the whole path between the small black hole state and the large black hole state and just need to consider the path from small (large) black hole state to the thermodynamic transition black hole state.

Based on the above analysis, we know that the trajectories from $r_m$ to $r_l$ and from $r_m$ to $r_s$ do not have a contribution, so we have the equations:
\be
\label{eq:20.22}
m_1=-m_2=-M_1,
\ee
\be
\label{eq:20.23}
m_3=-m_4=-M_2,
\ee
where the minus sign appearing in the expressions is associated with the presence of a turning point on the trajectory, namely a change from the instanton (anti-instanton) path to the anti-instanton (instanton) path~\cite{BK,AV}(See Appendix for the details). We call the $M_1$ as one pseudomolecule contribution to the probability and the $M_2$ as one anti-pseudomolecule contribution to the probability.

Then, the probability $P(r_s,t;r_s,t_0)$ and $P(r_l,t;r_s,t_0)$ can be given analytically. At first, one can calculate the probability $P(r_s,t;r_s,t_0)$. When there is zero pseudomolecule, the probability becomes $e^{-u(r_s)(t-t_0)}$, where $t-t_0$ represents the time interval staying at the small black hole state.

When there is one pseudomolecule, only an $a$ pseudomolecule contributes and the probability is
\be
\int_{t_0}^{\infty}dt_1 (-M_1)e^{-u(r_s)(t_1-t_0)}e^{-u(r_s)(t-t_1)},
\ee
where $(-M_1)$ is one $a$ pseudomolecule contribution to the probability, $t_1-t_0$ and $t-t_1$ represent the time interval staying at the small black hole state.

\begin{widetext}

When there are two pseudomolecules, two $a$ pseudomolecules contribute, or $b \rightarrow d$, the arrow represents the time sequence of the pseudomolecules. The probability is

\be
\ba
&\int_{t_0}^{\infty}dt_1\int_{t_1}^{\infty}dt_2(-M_1)^2 e^{-u(r_s)(t_1-t_0)} e^{-u(r_s)(t_2-t_1)}e^{-u(r_s)(t-t_2)}\\
&+\int_{t_0}^{\infty}dt_1\int_{t_1}^{\infty}dt_2(M_1M_2)e^{-u(r_s)(t_1-t_0)}e^{-u(r_l)(t_2-t_1)}e^{-u(r_s)(t-t_2)};
\ea
\ee
In the first term, $(-M_1)^2$ represents two $a$ pseudomolecule contribution to the probability, $t_1-t_0$, $t_2-t_1$ and $t-t_2$ are the time intervals staying at the small black hole state. In the second term, $(M_1M_2)$ represents one $b$ and one $d$ pseudomolecule contribution to the probability, $t_1-t_0$ and $t-t_2$ are the time intervals staying at the small black hole state, $t_2-t_1$ is the time interval staying at the large black hole state.

When there are three pseudomolecules, three $a$ pseudomolecules contribute, $a\rightarrow b\rightarrow d$, $b\rightarrow d\rightarrow a$, or $b\rightarrow c\rightarrow d$, the probability is

\be
\ba
&\int_{t_0}^{\infty}dt_1\int_{t_1}^{\infty}dt_2\int_{t_2}^{\infty}dt_3(-M_1)^3 e^{-u(r_s)(t_1-t_0)}e^{-u(r_s)(t_2-t_1)}e^{-u(r_s)(t_3-t_2)}e^{-u(r_s)(t-t_3)}\\
&+\int_{t_0}^{\infty}dt_1\int_{t_1}^{\infty}dt_2\int_{t_2}^{\infty}dt_3 (-M_1^2M_2) e^{-u(r_s)(t_1-t_0)}e^{-u(r_s)(t_2-t_1)}e^{-u(r_l)(t_3-t_2)}e^{-u(r_s)(t-t_3)}\\
&+\int_{t_0}^{\infty}dt_1\int_{t_1}^{\infty}dt_2\int_{t_2}^{\infty}dt_3(-M_1^2 M_2) e^{-u(r_s)(t_1-t_0)}e^{-u(r_l)(t_2-t_1)}e^{-u(r_s)(t_3-t_2)}e^{-u(r_s)(t-t_3)}\\
&+\int_{t_0}^{\infty}dt_1\int_{t_1}^{\infty}dt_2\int_{t_2}^{\infty}dt_3 (-M_1 M_2^2) e^{-u(r_s)(t_1-t_0)}e^{-u(r_l)(t_2-t_1)}e^{-u(r_l)(t_3-t_2)}e^{-u(r_s)(t-t_3)}.
\ea
\ee

In the long time limit, the probability $P(r_s,t;r_s,t_0)$ is the sum of the pseudomolecule number from $0$ to $\infty$; Thus, based on the condition $u(r_s)=u(r_l)=u(r_m)=u$, the probability is simplified as:
\be
\ba
\label{eq:15.10}
P(r_s,t;r_s,t_0)=&e^{-u(t-t_0)}-M_1\int_{t_0}^{\infty}dt_1 e^{-u(t_1-t_0)}e^{-u(t-t_1)}\\
&+M_1(M_1+M_2)\int_{t_0}^{\infty}dt_1\int_{t_1}^{\infty}dt_2 e^{-u(t_1-t_0)} e^{-u(t_2-t_1)} e^{-u(t-t_2)}\\
&-M_1(M_1+M_2)^2\int_{t_0}^{\infty}dt_1\int_{t_1}^{\infty}dt_2\int_{t_2}^{\infty}dt_3 e^{-u(t_1-t_0)}e^{-u(t_2-t_1)}e^{-u(t_3-t_2)}e^{-u(t-t_3)}\\
&+\dots\\
=&e^{-u(t-t_0)}+M_1\sum_{n=1}^{\infty}(-1)^n (M_1+M_2)^{n-1}\int_{t_0}^{\infty}dt_1\int_{t_1}^{\infty}dt_2\dots\int_{t_{n-1}}^{\infty}dt_n\\
&e^{-u(t_1-t_0)}e^{-u(t_2-t_1)}\dots e^{-u(t-t_n)},
\ea
\ee
where $t_0$=0.
\end{widetext}

By using the Laplace transform, we can obtain
\be
\ba
P(s)=\frac{1}{s+u}-\frac{M_1}{M_1+M_2}[\frac{1}{s+u}-\frac{1}{s+u+M_1+M_2}].
\ea
\ee

Inverting the Laplace transform, we can simplify the Eq.(\ref{eq:15.10}) as:
\be
\ba
P(r_s,t;r_s,0)=\frac{e^{-ut}}{M_1+M_2}[M_2+M_1e^{(-M_1-M_2)t}].
\ea
\ee

In our problem, $u(r_s)=u(r_m)=u(r_l)=u=0$, so
\be
\label{eq:3.70}
P(r_s,t;r_s,0)=\frac{1}{M_1+M_2}[M_2+M_1e^{(-M_1-M_2)t}].
\ee

An similar procedure can be applied to the calculation of $P(r_l,t;r_s,0)$, one can obtain:
\be
\label{eq:3.71}
P(r_l,t;r_s,0)=\frac{1}{M_1+M_2}[M_1-M_1e^{-(M_1+M_2)t}].
\ee

\subsection{The physical significance of one pseudomolecule or one anti-pseudomolecule contribution to the probability and the kinetic rates}

We can consider a model for a particle moving in a double well potential with the two stable states denoted as $A$ and $B$. We assume that the transition rate from state $A$ to state $B$ is $k_A$ and the transition rate from state $B$ to state $A$ is $k_B$, while the initial state of the particle is at state $A$. The $P_A(\tau)$ represents the probability of the particle staying at the state $A$ at time $\tau$, and the $P_B(\tau)$ represents the probability of the particle staying at the state $B$ at time $\tau$. Then, one can write the classical master equation as:
\be
\label{eq:15.20}
\frac{d P_A(\tau)}{d\tau}=-k_A P_A(\tau)+k_B P_B(\tau).
\ee

The total probability should be conserved, we can obtain
\be
\label{eq:15.21}
P_A(\tau)+P_B(\tau)=1.
\ee

When substituting eq.(\ref{eq:15.21}) back to eq.(\ref{eq:15.20}) and integrating the $\tau$ from $0$ to $t$, we can obtain:
\be
\label{eq:3.72}
P_A(t)=\frac{1}{k_A+k_B}[k_B+k_Ae^{-(k_A+k_B)t}].
\ee

The $P_B(t)$ is given by
\be
\ba
\label{eq:3.73}
P_B(t)&=1-P_A(t)\\
&=\frac{1}{k_A+k_B}[k_A-k_Ae^{-(k_A+k_B)t}].
\ea
\ee

In the small-large black hole phase transition, the Gibbs free energy landscape has the double well shape as shown in Fig.~\ref{fgFE}. Eq.(\ref{eq:3.72}) and Eq.(\ref{eq:3.73}) can then be used to describe the time evolution of the transition probability during the phase transition, and we should obtain the same results as Eq.(\ref{eq:3.70}) and Eq.(\ref{eq:3.71}) after taking the state $A$ and $B$ as the small and large black hole state respectively. When we compare the equation (\ref{eq:3.70}) and (\ref{eq:3.71}) to the equation (\ref{eq:3.72}) and (\ref{eq:3.73}), the physical significance of $M_1$ and $M_2$ can be easily seen: The $M_1$ represents the transition rate from the small black hole to large black hole, and the $M_2$ represents the transition rate from the large black hole to the small black hole.

Furthermore, based on the equation (\ref{eq:3.70}) and (\ref{eq:3.71}), the total kinetic rate is given by
\be
\label{eq:4.70}
k=M_1+M_2,
\ee
which determines the rate or the time scale (inverse of the rate $\frac{1}{k}$) of the probability evolution for $P(r_s,t;r_s,0)$ and $P(r_l,t;r_s,0)$.

\subsection{The second order effects}

The fluctuation effects on the dominate path can be considered. Then the phase transition rate and the probability evolution will be modified.

We replace $D\beta G(r)$ by $U(r)$ in Eq.(\ref{eq:3.1}) for simplification, and the probability is given by:
\be
\ba
P(r_t,t_1;r_0,t_0)=&\exp[-\frac{[U(r_t)-U(r_0)]}{2D}]\int_{r_0}^{r_t}Dr\exp\lbrace\\
&-\frac{1}{D}\int_{t_0}^{t_1}dt[\frac{(\dot{r})^2}{4}+\frac{(U'(r))^2}{4}-\frac{D}{2}U''(r)]\rbrace\\
=&\exp[-\frac{[U(r_t)-U(r_0)]}{2D}]\\
&\times\int_{r_0}^{r_t}Dr \exp[ -\frac{S(r(t))}{D}]\\
\ea
\ee

We denote $\int_{r_0}^{r_t}Dr \exp[ -\frac{S(r(t))}{D}]$ as $K(r_t,t_1;r_0,t_0)$. The action can be expanded around the classical path to the second order variation in $y(t)=r(t)-r_{cl}(t)$, one yields:
\be
\ba
\label{eq:1.90}
K(r_t,t_1;r_0,t_0)=&\exp\{-\frac{S[r_{cl}(t)]}{D}\}\int Dy(t)\exp\{-\frac{1}{2D}\\
&\times\int_{t_0}^{t_1}y(t)[-\frac{1}{2}\frac{d^2}{dt^2}+V''(r_{cl}(t))]y(t)dt\},
\ea
\ee
where $V(r)=\frac{(U'(r))^2}{4}-\frac{D}{2}U''(r)$.

We expand y(t) on an infinite orthogonal basis $\{y_n(t)\}$ which are also the eigenfunctions of $-\frac{1}{2}\frac{d^2}{dt^2}+V''(r_{cl}(t))$, these eigenfunctions satisfy the equation (\ref{eq:1.93}), (\ref{eq:1.94}), (\ref{eq:1.95}) and (\ref{eq:1.96}). By using the Gauss integral, the equation (\ref{eq:1.90}) becomes~\cite{BF,BB,BH}:
\be
\ba
K(r_t,t_1;r_0,t_0)&=\frac{N}{\det[-\frac{1}{2}\frac{d^2}{dt^2}+V''(r_{cl}(t))]}\exp[-\frac{S(r_{cl}(t))}{D}]\\
&=\frac{N}{\sqrt{\prod_n \lambda_n}}\exp[-\frac{S(r_{cl}(t))}{D}],
\ea
\ee
where $\lambda_n$ are the eigenvalues of the operator $-\frac{1}{2}\frac{d^2}{dt^2}+V''(r_{cl}(t))$, $N$ is a constant. More descriptions are given in the appendix

This equation holds for monostable potential, but will break down for the potential in our case which always has a eigenfunction $\dot r_{cl}(t)$ with zero eigenvalue. Thus, the Gaussian approximation of the corresponding fluctuation modes will break down. These modes are called zero modes and their physical origin is the time translational invariance of the system~\cite{BB,BF}. Considering the case of one zero mode (or one instanton), the equation (\ref{eq:1.90}) can be replaced by~\cite{AV,AW,BF,BH}:

\be
\ba
K(r_t,t_1;r_0,t_0)=&\int_{t_0}^{t_1}d\tau_0 \sqrt{\frac{\lambda_0}{4\pi D\psi_{\lambda_0}(t_1)}} \sqrt{\frac{S(r_{cl})}{4\pi D}}\\
&\times\exp[-\frac{S(r_{cl})}{D}],
\ea
\ee
where $\lambda_0$ and $\psi_{\lambda_0}(t)$ satisfy the equation (\ref{eq:6.10}), (\ref{eq:6.11}) and (\ref{eq:6.12}), and $\sqrt{\frac{S(r_{cl})}{4\pi D}}$ is the integration measure of the variables $\tau_0$.

The path integral problems of the second order in the symmetric double well have been explored by~\cite{BB,BC,BD,BE,BF,BG,BH}, while the problems become quite difficult in the asymmetric double well. Because of the asymmetry, the instanton has the different asymptotic behaviours in the two sides of the time axis, and the calculations become tedious. This has been explored in~\cite{AV,AW}, which is based on the method of~\cite{BF,BG,BH}. The main procedures are shown in the appendix. Assume that there are no interactions between these instantons, then the dilute gas approximation can be used to obtain the final probability by summing over the multi-instantons. After a heavy algebra, the final results show the probabilities driven by the classical paths have a correction as~\cite{AV,AW}:

\be
P(r_s,t;r_s,0)=\sqrt{\frac{\beta G''(r_s)}{2\pi}}\frac{1}{M_1+M_2}[M_2+M_1e^{(-M_1-M_2)t}],
\ee
\be
P(r_l,t;r_s,0)=\sqrt{\frac{\beta G''(r_l)}{2\pi}}\frac{1}{M_1+M_2}[M_1-M_1e^{-(M_1+M_2)t}],
\ee
where $M_1$ and $M_2$ have a correction as:
\be
\label{eq:10.1}
M_1 \rightarrow \frac{\beta D\sqrt{|G''(r_m)|G''(r_s)}}{2\pi}*M_1,
\ee
\be
\label{eq:10.2}
M_2 \rightarrow \frac{\beta D\sqrt{|G''(r_m)|G''(r_l)}}{2\pi}*M_2.
\ee

\subsection{The numerical results}

\begin{figure}[tbp]
\centering
\includegraphics[width=0.48\textwidth]{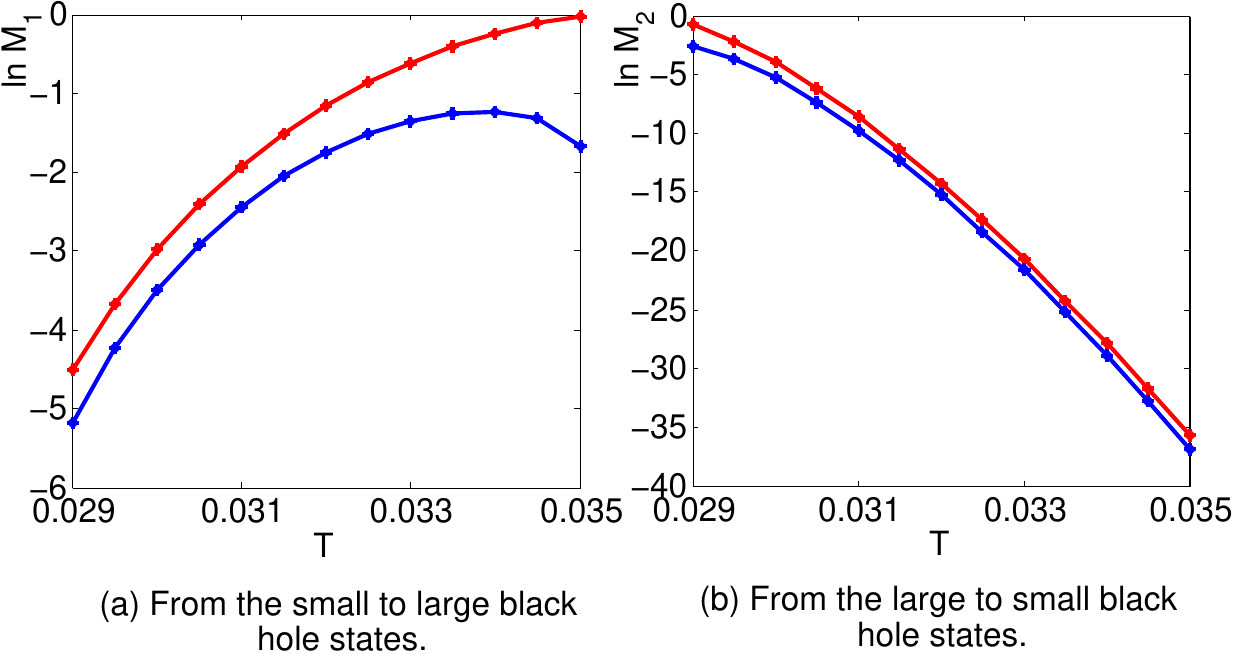}
\caption{The kinetic rate of the phase transition between the small and large black hole states at the low diffusion coefficient limit. The horizontal axis is the temperature, and the vertical coordinate is the logarithm of transition rate. The red lines represent only the zero order effects are considered, while the blue lines represent both the zero order and second order effects are considered.}
\label{fgPTR}
\end{figure}

The phase transition rate is an important entity in the dynamics of phase transition process, which quantifies the time scale of the small (large) black hole state switching to the large (small) black hole state. Based on the equation (\ref{eq:3.50}), we use the classical pseudomolecule paths to obtain the temperature dependence of the phase transition rates in Fig.~\ref{fgPTR} (red lines). If we take into account of the second order effects, there are some corrections to the phase transition rates as shown in eq.~(\ref{eq:10.1}) and eq.~(\ref{eq:10.2}). We also plot the temperature dependence of the phase transition rates including the second order effects in Fig.~\ref{fgPTR} (blue lines). Note that the vertical coordinate is the logarithm of the phase transition rate.

When we analyze the phase transition rate without the second order effects (red lines). The results show, upon the increase of the temperature, the kinetic rate of phase transition will increase from the small to large black hole states transition and decrease from the large to small black hole states transition. As shown in Fig.~\ref{fgFE}, we  know that the barrier height from the small black hole state to the large black hole state through the thermodynamic transition black hole state will decrease with the temperature. This indicates that the small black hole state should be easier to be switched to the large black hole state as the temperature increases. The barrier height from the large to the small black hole states through the thermodynamic transition black hole state will increase with the temperature. Then the large black hole state should be more difficult to be switched to the small black hole state. These are consistent with the picture of quantified rates in Fig.~\ref{fgPTR}. Furthermore, when the Gibbs free energies of small and large black holes are equal ($T=0.0298$), the phase transition rates of $M_1$ and $M_2$ should be equal. This also corresponds to our resulting rates well.

When we analyze the phase transition rate including the second order effects (blue lines), we can see that both these curves are near to the corresponding curves without the second order effects (red lines). Furthermore, we note that the curve in the left panel has a inflection point comparing with the curve without the second order effects. It is an interesting phenomenon which means that the second order effects can become significant compared to the zero order effects when the temperature is high. After taking into account of the second order effects, the phase transition rate should be determined by both the barrier height of the free energy landscape in the exponential and the second derivatives of the free energy landscape at the basin and at the barrier (saddle) in the prefactor. When the temperature is high, the barrier height of the free energy landscape from the small black hole to the large black hole through the thermodynamic transition black hole does not change significantly upon the increase of the temperature. However, the prefactor or second order derivatives of the free energy landscape at both the small black hole basin and the thermodynamic transition black hole barrier or saddle decrease clearly. Thus, the second order effects become more important than the zero order effects, and the rate of phase transition decreases accordingly. In the right panel, we can not observe such an inflection point because of the obvious variation of the barrier height.

\begin{figure}[t]
\centering
\includegraphics[width=0.48\textwidth]{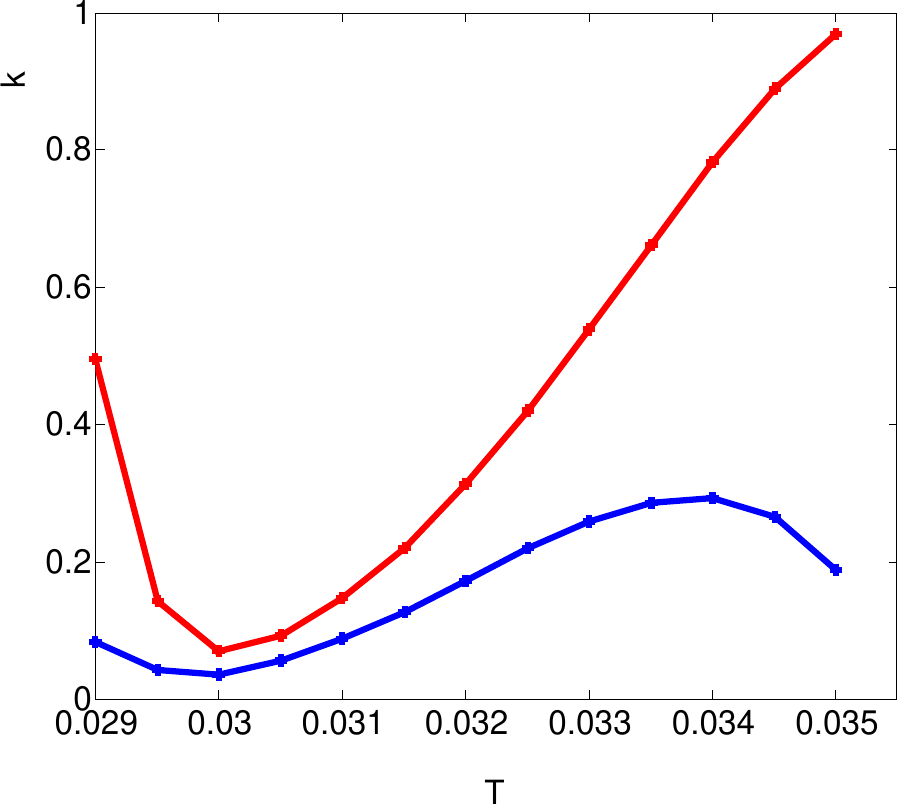}
\caption{The temperature dependence of the total kinetic rate. The red lines represent only the zero order effects are considered, while the blue lines represent both the zero order and second order effects are considered.}
\label{fgTPTR}
\end{figure}

The temperature dependence of the total kinetic rates in both the cases of zero order (red lines) and of the one including the second order (blue lines) have been plotted in Fig.~\ref{fgTPTR}. It can be seen that the total kinetic rate decreases with the temperature at first and then increases with the temperature in the zero order effects. The total kinetic rate indicates the time scale of the probability evolution, which is the combination of the rates from small to large and from large to small black hole phase transition. The two kinetic behaviors with the temperature reflect the temperature dependence of each individual transition rate. When the temperature is high, the rate from the small to large black hole state dominates in the total kinetic rate, and the inflection point of the total kinetic rate will appear after taking into account the second order effects for the same reason as the left panel in Fig.~\ref{fgPTR}. When the temperature is low, the rate of the transition from large black hole to small black hole dominates in the total kinetic rates.

\begin{figure}[h]
\centering
\includegraphics[width=0.48\textwidth]{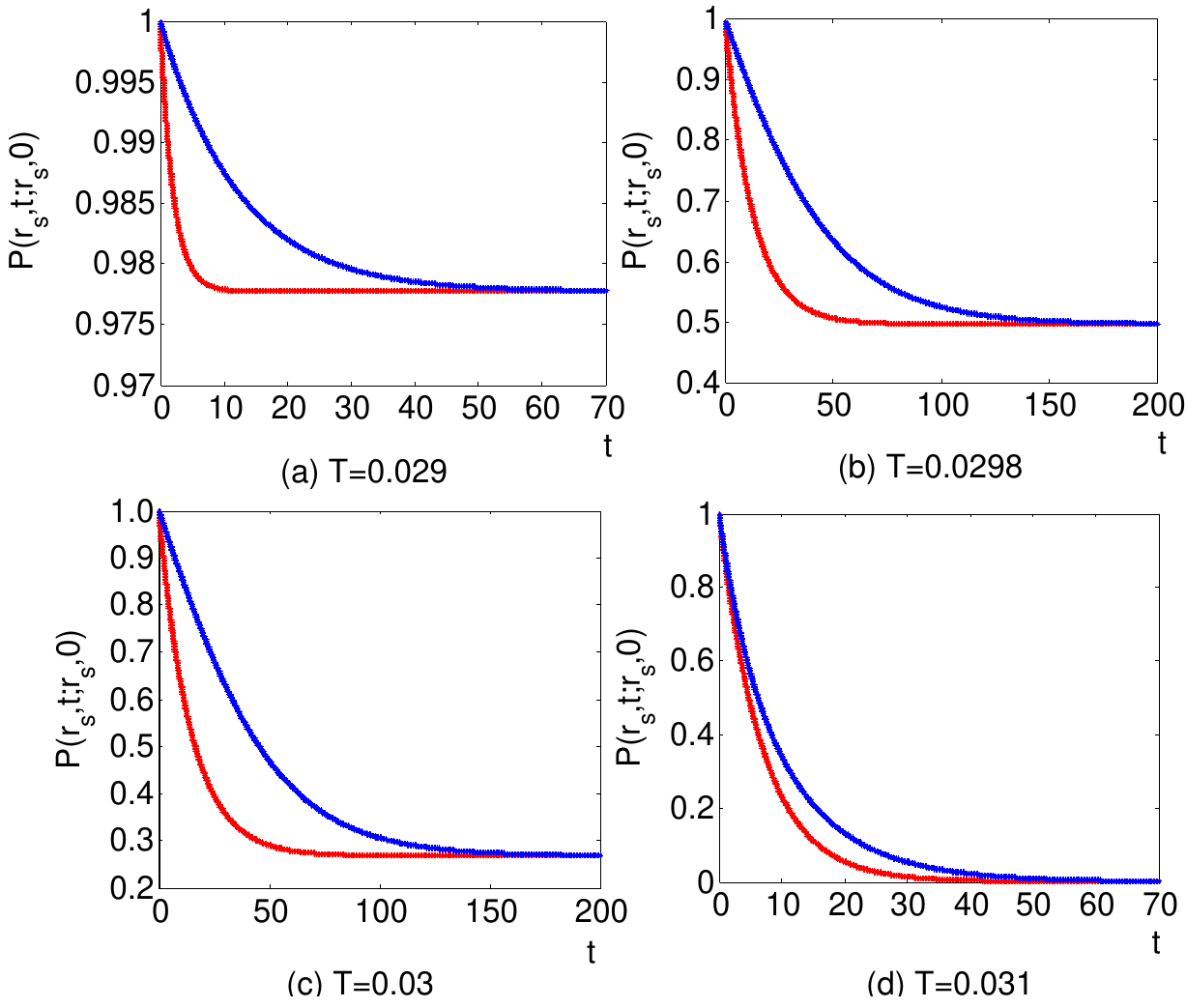}
\caption{The time evolution of the probability $P(r_s,t;r_s,0)$ at different temperatures $T=0.029$, $0.0298$, $0.03$ and $0.031$. The red lines represent only the zero order effects are considered, while the blue lines represent both the zero order and second order effects are considered.}
\label{fgSSP}
\end{figure}

\begin{figure}[h]
\centering
\includegraphics[width=0.48\textwidth]{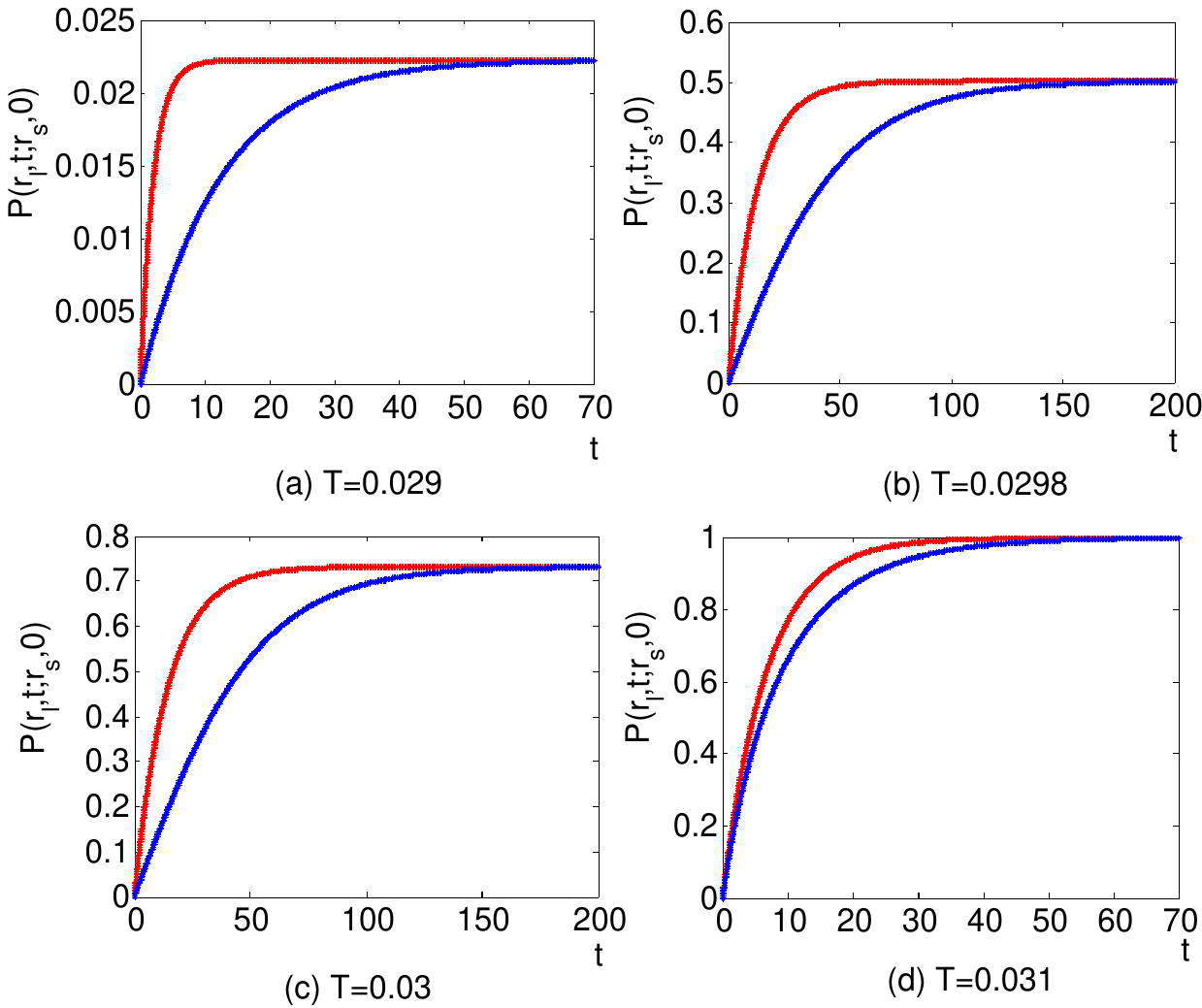}
\caption{The time evolution of the probability $P(r_l,t;r_s,0)$ at different temperatures $T=0.029$, $0.0298$, $0.03$ and $0.031$. The red lines represent only the zero order effects are considered, while the blue lines represent both the zero order and second order effects are considered.}
\label{fgLSP}
\end{figure}

Furthermore, the time evolutions of $P(r_s,t;r_s,0)$ and $P(r_l,t;r_s,0)$ at different temperatures in both two cases are given in Fig.~\ref{fgSSP} and Fig.~\ref{fgLSP}. As seen, the $P(r_s,t;r_s,0)$ and $P(r_l,t;r_s,0)$ become steady when $t$ is large, the time of the probability being steady is determined by the total kinetic rate in Fig.~\ref{fgTPTR}.

The stationary probability can reflect the thermodynamic stability, and it is determined by the value of the Gibbs free energy via the Boltzmann distribution. Whether we take into account of the second order effects or not, the stationary probability should be equal. As shown in Fig.~\ref{fgSSP} and Fig.~\ref{fgLSP}, the red line (without the second order effects) and the blue line (with the second order effects) are steady to the same value at the same temperature. When we analyze the Fig.~\ref{fgSSP} and Fig.~\ref{fgLSP} separately, we can see that the steady state probability $P(r_s,t;r_s,0)$ decreases and the steady state probability $P(r_l,t;r_s,0)$ increases as the temperature increases. As shown in Fig.~\ref{fgFE}, when the temperature increases, the free energy of the small black hole state decreases slower than that of the large black hole state. The Boltzmann distribution tells us that the steady state probability $P(r_s,t;r_s,0)$ decreases and the steady state probability $P(r_l,t;r_s,0)$ increases with the temperature. When we compare the Fig.~\ref{fgSSP} with the Fig.~\ref{fgLSP}, we can perform the following analyses. At $T=0.0298$, the Gibbs free energies of the small black hole state and the large black hole state are equal, the steady state probability $P(r_s,t;r_s,0)$ and $P(r_l,t;r_s,0)$ should be equal. As shown in Fig.~\ref{fgSSP} and Fig.~\ref{fgLSP}, they are both equal to $0.5$. When $T=0.029<0.0298$, the free energy of the small black hole state is lower than the free energy of the large black hole state. This indicates that the small black hole state is more stable, thus the steady state probability $P(r_s,t;r_s.0)$ is higher than the steady probability $P(r_l,t;r_s,0)$. When $T=0.03$ and $0.031$, they are both larger than $T=0.0298$. Then the large black hole state has lower free energy and thus becomes more stable. Correspondingly, the steady state probability $P(r_l,t;r_s,t)$ is higher than the steady state probability $P(r_s,t;r_s,0)$.

\section{Conclusions}
\label{Conclusions}
In conclusion, we have formulated a path integral framework to investigate the dynamical phase transition of RNAdS black hole under the free energy landscape. There are three macroscopic emergent phases in the extended phase space. The small and large black hole states are stable and the thermodynamic transition black hole state is unstable. Under the thermal fluctuations, the phase transition is possible between the small and large black hole states. The corresponding dynamics can be described by the stochastic Langevin equation, where the thermodynamic driving force is provided by the underlying Gibbs free energy and the stochastic force comes from the thermal fluctuations. The contributions of the different paths to the weights or probabilities are on the exponentials. Thus, the dominant path, which satisfies the Euler-Lagrangian equation due to the maximization of the weights or minimization of the action, can be regarded as the main path in which the phase transition proceeds. Based on the dominant path, we derive the analytical formula for the time evolution of the transition probability. After comparing with an analogous model for a particle moving in the double well potential, we find that the contribution to the probability from one ``pseudomolecule" (``anti-pseudomolecule") can be interpreted as the phase transition rate from the small (large) to large (small) black hole state. The numerical results show a good consistency with the underlying free energy landscape topography. This work provides a new framework to investigate the dynamics of black hole phase transition, which can address the important issues of both the kinetic path and the phase transition rate. This framework can also be used to investigate other kinds of black hole phase transition dynamics.

\section*{Acknowledgments}
C. H. Liu thanks Hong Wang and He Wang for the helpful discussions. C. H. Liu thanks the support in part by the National Natural Science Foundation of China Grants 21721003.

\vskip 1cm
\appendix

\section{The second order effects}

The Eq.(\ref{eq:1.90}) is written again as:
\be
\ba
\label{eq:3.90}
&K(r_t,t_1;r_0,t_0)=e^{-\frac{S[r_{cl}(t)]}{D}}\int Dy(t)\\
&\exp\{-\frac{1}{2D}\int_{t_0}^{t_1}dt y(t)[-\frac{1}{2}\frac{d^2}{dt^2}+V''(r_{cl}(t))]y(t)\}
\ea
\ee

One can expand y(t) on an infinite orthogonal basis $\{y_n(t)\}$ which is also the eigenfunction of the second variational derivative of S~\cite{BF}:
\be
\label{eq:1.93}
y(t)=\sum_n c_n y_n(t),
\ee
\be
\label{eq:1.94}
[-\frac{1}{2}\frac{d^2}{dt^2}+V''(r_{cl}(t))]y_n(t)=\lambda_n y_n(t),
\ee
\be
\label{eq:1.95}
y_n(t_0)=y_n(t_1)=0,
\ee
\be
\label{eq:1.96}
\int_{t_0}^{t_1}y_n(t)y_m(t) dt=\delta_{mn}.
\ee

By using the Gauss integral, the equation (\ref{eq:3.90}) becomes
\be
\ba
\label{eq:7.75}
K(r_t,t_1;r_0,t_0)=&\frac{N}{\det[-\frac{1}{2}\frac{d^2}{dt^2}+V''(r_{cl}(t))]}\exp[-\frac{S(r_{cl}(t))}{D}]\\
=&\frac{N}{\det[-\frac{1}{2}\frac{d^2}{dt^2}+\frac{1}{2}w^2]}\frac{\det[-\frac{1}{2}\frac{d^2}{dt^2}+\frac{1}{2}w^2]}{\det[-\frac{1}{2}\frac{d^2}{dt^2}+V''(r_{cl}(t))]}\\
&\times\exp[-\frac{S(r_{cl}(t))}{D}]\\
=&\frac{N}{\sqrt{\prod_n\lambda_n^{(h)}}}\frac{\sqrt{\prod_n\lambda_n^{(h)}}}{\sqrt{\prod_n \lambda_n}}\exp[-\frac{S(r_{cl}(t))}{D}].
\ea
\ee
Here we have brought in the well-known harmonic solution $\frac{N}{\sqrt{\prod_n\lambda_n^{(h)}}}$ to eliminate the constant $N$, then the key issue becomes how to calculate the factor $\frac{\sqrt{\prod_n\lambda_n^{(h)}}}{\sqrt{\prod_n \lambda_n}}$~\cite{BB,BF}.

Based on the same zero points and the same pole points at the two sides of the follow equation, it was proved that~\cite{BF,BG,BH}
\be
\label{eq:20.10}
\det[\frac{-\frac{1}{2}\partial_t^2+V^{(1)}-\lambda}{-\frac{1}{2}\partial_t^2+V^{(2)}-\lambda}]=\frac{\psi_{\lambda}^{(1)}(T/2)}{\psi_{\lambda}^{(2)}(T/2)},
\ee
where $\psi_{\lambda}(t)$ is the corresponding solution satisfying:
\be
\label{eq:6.10}
(-\frac{1}{2}\partial_t^2+V^{(i)})\psi_{\lambda}^{(i)}(t)=\lambda\psi_{\lambda}^{(i)}(t),
\ee
\be
\ba
\label{eq:6.11}
\psi_{\lambda}^{(i)}(t_0)=0,\quad   \partial_t\psi_{\lambda}^{(i)}(t_0)=1,
\ea
\ee
where $i=1$, $2$.

The operator $-\frac{1}{2}\partial_t^2+V^{(i)}$ has an eigenvalue $\lambda_n$, only if
\be
\label{eq:6.12}
\psi_{\lambda_n}^{(i)}(t_1)=0.
\ee

Taking $\lambda=0$ in Eq.(\ref{eq:20.10}), the problem is then changed to evaluate the ratio of the corresponding lowest eigenfunction. This condition holds for monostable potentials. However, there is always an eigenfunction $\dot r_{cl}(t)$ with zero eigenvalue in our case, and the Gaussian approximation of the corresponding fluctuation modes will break down. These modes are called zero modes~\cite{BF,BG,BH,AV,AW}. The eq.(\ref{eq:7.75}) should factor out the zero modes and evaluate the determinant with zero eigenvalue omitted. When considering that there is only one instanton (or one zero mode), one can rewrite Eq.(\ref{eq:7.75}) as:
\be
\ba
K(r_t,t_1;r_0,t_0)=&\frac{N\sqrt{\psi_{\lambda_0}^{(h)}(t_1)}}{\sqrt{\prod_n\lambda_n^{(h)}}}\sqrt{\frac{\lambda_0}{\psi_{\lambda_0}(t_1)}}\int dc_0\\
&\times\exp[-\frac{S(r_{cl}(t))}{D}].
\ea
\ee

Based on the time invariance of the instantons, one can obtain:
\be
\ba
\delta(r(t+\tau_0))&=\frac{dr(t)}{dt}\delta \tau_0=y_0(t)\delta c_0\\
&=(\frac{S_{cl}}{m})^{-\frac{1}{2}}\dot r_{cl}(t)\delta c_0.
\ea
\ee

Then, one can replace the $dc_0$ integration by an integration over the position of the center of the instanton $d\tau_0$:
\be
dc_0=\sqrt{\frac{S_{cl}}{m}}d\tau_0.
\ee

The $K(r_t,t_1;r_0,t_0)$ then becomes:
\be
\ba
K(r_t,t_1;r_0,t_0)=&\int_{t_0}^{t_1}d\tau_0 \sqrt{\frac{\lambda_0}{4\pi D\psi_{\lambda_0}(t_1)}} \sqrt{\frac{S(r_{cl})}{4\pi D}}\\
&\times\exp[-\frac{S(r_{cl})}{D}],
\ea
\ee
and the task changes to evaluate the $\frac{\lambda_0}{\psi_{\lambda_0}(t_1)}$.

Based on the known solution of eq.~(\ref{eq:6.10}) with zero eigenvalue $x_1(t)\propto\dot r_{cl}(t)$, one can find another solution $y_1(t)$ with zero eigenvalue by the D'Alembert's construction~\cite{BB,BF}:
\be
\ba
\label{eq:9.90}
y_1(t)=Wx_1(t)\int^t\frac{dt'}{x_1^2(t')}.
\ea
\ee

Taking the derivative with respect to the time, one can obtain
\be
W=x_1(t)\dot y_1(t)-y_1(t)\dot x_1(t),
\ee
where $W$ is actually the Wronskian determinant.

After using the classical equation of motion, one can obtain the asymptotic expression of the instanton solution $\dot r_{cl}(t)$ when $t\ll \tau_0$ and $t\gg \tau_0$(the detailed derivation can be seen in ~\cite{AV}). Then the asymptotic expression of $x_1(t)$ is driven by the equation
\be
x_1(t)=\sqrt{\frac{m}{S_{cl}}}\dot r_{cl}(t),
\ee
where $\sqrt{\frac{m}{S_{cl}}}$ is the normalized factor. The asymptotic expression of $y_1(t)$ when $t\ll \tau_0$ and $t\gg \tau_0$ can be given based on the D'Alembert's construction (\ref{eq:9.90}). Thus, the function $\psi_{\lambda_0}(t)$ satisfying the equation (\ref{eq:6.10}), (\ref{eq:6.11}) and (\ref{eq:6.12}) can be given by the linear combination of $x_1(t)$ and $y_1(t)$~\cite{AV}. By transforming the equation (\ref{eq:6.10}) into the integral equation and iterating once, one can obtain~\cite{BF,AV}:
\be
\psi_{\lambda_0}(t)=\psi(t)-\frac{2\lambda_0}{W}\int_{t_0}^{t}dt'[y_1(t)x_1(t')-x_1(t)y_1(t')]\psi_{\lambda_0}(t'),
\ee
where $\psi_{\lambda_0}(t_0)=0$.

One can take $t=t_1$ and use the equation (\ref{eq:6.12}), the ratio of $\frac{\lambda_0}{\psi(t_1)}$ can be obtained:
\be
\frac{\lambda_0}{\psi(t_1)}=\frac{W}{2}\{\int_{t_0}^{t_1}dt'[y_1(t_1)x_1(t')-x_1(t_1)y_1(t')]\psi_{\lambda_0}(t')\}^{-1}.
\ee

Based on the known asymptotic expression of $x_1(t)$ and $y_1(t)$, the $\frac{\lambda_0}{\psi(t_1)}$ can be calculated analytically. Actually, the result of $\frac{\lambda_0}{\psi(t_1)}$ is proportional to $\dot r_{cl}(t_0)\dot r_{cl}(t_1)$, where $t_0$ and $t_1$ are the end points of the instanton path. For a pseudomolecule composed by an instanton and an anti-instanton, there is a turning point of the path, namely a sign change of $\dot r_{cl}$, which will result in the minus sign in eq.~(\ref{eq:20.22}) and eq.~(\ref{eq:20.23})~\cite{BK,AV}.

Taking into account all the above terms, one can calculate the corresponding $K(r_t,t_1;r_0,t_0)$ which only has one instanton. For the multi-instantons, the probability can be summed by the dilute gas approximation. After the heavy algebra, the results are finally given as the following~\cite{AV}:

\be
P(r_s,t;r_s,0)=\sqrt{\frac{\beta G''(r_s)}{2\pi}}\frac{1}{M_1+M_2}[M_2+M_1e^{(-M_1-M_2)t}],
\ee
\be
P(r_l,t;r_s,0)=\sqrt{\frac{\beta G''(r_l)}{2\pi}}\frac{1}{M_1+M_2}[M_1-M_1e^{-(M_1+M_2)t}],
\ee
where $M_1$ and $M_2$ have a correction as:
\be
\label{eq:10.8}
M_1 \rightarrow \frac{\beta D\sqrt{|G''(r_m)|G''(r_s)}}{2\pi}*M_1,
\ee
\be
\label{eq:10.9}
M_2 \rightarrow \frac{\beta D\sqrt{|G''(r_m)|G''(r_l)}}{2\pi}*M_2.
\ee

\end{document}